\documentclass[prl,twocolumn,aps,amsmath,amssymb,nofootinbib]{revtex4-1}
\usepackage{hyperref}
\usepackage{bm}
\usepackage{amsmath}
\usepackage{amsfonts}

\usepackage{amssymb}

\newcommand{\bg}{ \begin{gather} }
\newcommand{\eg}{\end{gather}}
\newcommand{\be}{ \begin{equation} }
\newcommand{\ee}{\end{equation}}
\newcommand{\bbea}{ \begin{eqnarray} }
\newcommand{\eeea}{\end{eqnarray}}

\def \be {\begin{equation}}
\def \ee {\end{equation}}
\def \bea {\begin{align}}
\def \eea {\end{align}}
\def \p {\partial}

\def \BC {\begin{cases}}
\def \EC {\end{cases}}

\usepackage{graphicx}
\usepackage{dcolumn}
\usepackage{bm}
\usepackage{verbatim}
\usepackage{color}
\usepackage{ulem}
\usepackage{epsf}
\usepackage{mathtext} 				

\begin{document}

\title{Carbon nanotubes for polarization sensitive terahertz plasmonic interferometry}

\author{Y. Matyushkin$^{1,2,3,4}$, S. Danilov$^2$, M.  Moskotin$^{2,3}$, G. Fedorov$^{2,3}$, A. Bochin$^{5}$, I. Gorbenko$^{5}$, V. Kachorovskii$^{5,6}$, and S. Ganichev$^{1,6}$}

\affiliation{$^1$National Research University Higher School of Economics, Moscow 101000 Russia}
\affiliation{$^2$Terahertz Center, University of Regensburg, Regensburg 93040, Germany}
\affiliation{$^3$Moscow Institute of Physics and Technology National Research University, 141700	Dolgoprudny, Russia}
\affiliation{$^4$Physics Department, Moscow Pedagogical State University (MPSU), 119435 Moscow, Russia}
\affiliation{$^5$Ioffe Institute, 194021 St. Petersburg, Russia}
\affiliation{$^6$CENTERA, Institute of High Pressure Physics PAS, 01142 Warsaw, Poland}



\begin{abstract}
 We report on helicity sensitive photovoltaic  terahertz radiation response of a carbon nanotube made in a configuration of a field effect transistor. We find that the magnitude of the rectified voltage is different for clockwise and anticlockwise circularly polarized radiation. We demonstrate that this effect is a fingerprint of the plasma waves interference in the transistor  channel. We also find that that the presence of the helicity- and phase-sensitive interference part of the photovoltaic response is a universal phenomenon which is obtained in the systems of different dimensionality with different single particle spectrum. Its magnitude is characteristic of the plasma wave decay length. This opens up a wide avenue for applications in the area of plasmonic interferometry.
\end{abstract}

\maketitle
\section{Introduction}

Plasmonics  is a rapidly growing area of research with a huge potential in terahertz photovoltaics ~\cite{Mittelman2013,Dhillon2017}. In particular, plasmonic effects  \cite{Dyakonov1993,Dyakonov1996} were already very successfully used for robust, gate-voltage-tunable, and highly sensitive room temperature THz photodetectors based on field-effect transistors (FET), see e.g. \cite{Vicarelli2012,Koppens2014,Bandurin2018}. Recent impressive progress in plasmonics is partially associated with appearance of novel highly-conducting carbon nanomaterials, such as graphene ~\cite{Grigorenko2012,DiPietro2013,Koppens2014,Hartmann2014, Giorgianni2016,Autore2017,Politano2017,Yang2018,Bandurin2018,Agarwal2018} and carbon nanotubes ~\cite{He2015,Ryzhii2016,Wang2018review}. Record mobilities ~\cite{Laird2015} observed in these materials are very promising for fabrication of plasmonic devices with high quality factors. Very recently, an extremely attractive opportunity for a further breakthrough in the area of plasmonics showed up: it has been predicted ~\cite{Drexler2012,Romanov2013, Gorbenko2018, Gorbenko2019}  that a conventional field-effect transistor (FET) can operate as a tunable plasmonic interferometer. This idea opens up an opportunity for applications in phase-sensitive photovoltaics. A key ingredient of a tunable plasmonic interferometer is so called phase asymmetry \cite{Gorbenko2018,Gorbenko2019}  that can be introduced via a simple modification to the antenna design. Such an asymmetry allows one for conversion of interfering plasma waves into dc signal due to different types of rectification effects, for example, due to  hydrodynamic nonlinear phenomena.    This concept was successfully implemented in our recent work ~\cite{Matyushkin2020} using a graphene-based FET. Gate-voltage-tunability, as well as the frequency-and phase- sensitivity of this graphene-based interferometer was demonstrated and successfully  interpreted  by using  hydrodynamic approach. 

Previous works ~\cite{Drexler2012,Romanov2013, Gorbenko2018, Gorbenko2019,Matyushkin2020} showed that plasmonic interferometer can be realized based on 2D systems.  
In this paper, we demonstrate both experimentally and theoretically that quasi-one-dimensional and one-dimensional systems like carbon nanotubes that are very promising for plasmonic applications can also effectively operate as plasmonic interferometers.   
Our main experimental result is that the magnitude of the rectified voltage is different for clockwise and anticlockwise circularly polarized radiation. Its interpretation is provided within a generalized approach taking into account plasmonic nonlinearities. We find that that the presence of the helicity- and phase-sensitive interference part of the photovoltaic response is a universal phenomenon which is (up to unimportant numerical coefficients) is obtained in the systems of different dimensionality with various single particle spectrum. Its magnitude is characteristic of the plasma wave decay length. This creates new options for applications in the area of plasmonic interferometry.

	\begin{figure}[htbp]
	\centering\includegraphics[width=7cm]{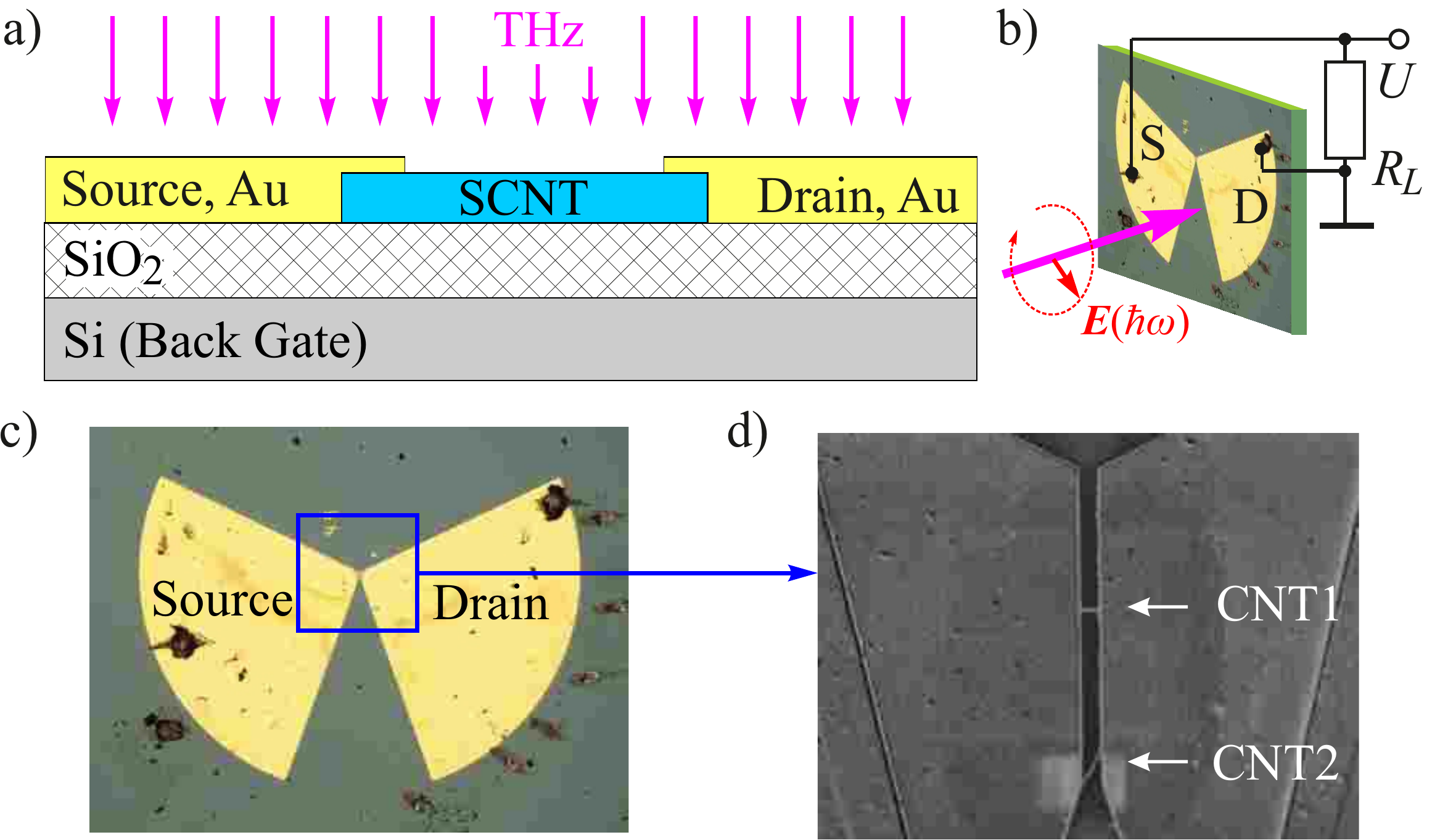}
	\caption {Devices configuration and characterization. (a) Structures cross-section. (b) Experimental set-up. (c) and (d)  Optical and scanning electron microscopy (SEM) images illustrating the device layout with source and drain electrodes connected to sleeves of a bent bow-tie antenna.
	}
	\label{characterization}
\end{figure}

\section{Samples and methods}

Schematic cross-section of the investigated FET is shown in  Fig.~\ref{characterization}(a) and Fig.~\ref{characterization}(c,d) shows optical photo and SEM-picture respectively.  Carbon nanotubes were synthesized using chemical vapor deposition (CVD) technique with H$_2$ and CH$_4$ precursors. Catalytic nanoparticles  (Fe(NO$_3$)$_3$; MoO$_2$ and Al2O$_3$) were spin-coated on oxidized silicon substrate from isopropyl alcohol suspension. Such catalytic nanoparticles let us get carbon nanotubes with diameters from 2 to 4 nm - single walled and doubled CNTs mostly ~\cite{Kong1998,Chen2007}. Contact interfaces to SWCNT and THz bow-tie antenna, were fabricated in two different parts of technological route with help e-beam lithography and lift-off technology. SWNTS contact interfaces were made from Au (25 nm) by e-beam sputtering. To realize the helicity-sensitive terahertz plasmonic interferometer, the antenna sleeves were bent by $40^\circ$ as shown in Figure~\ref{characterization}(c) and used as the source and drain electrodes. THz antennas for the radiation coupling with CNT were fabricated with thermal vacuum evaporation (Ti-Au 10-200~nm). At the final fabrication step all the CNTs outside the channel area (in the Fig.~\ref{characterization}(d)) were etched in an oxygen plasma. As seen from the Fig.~\ref{characterization}(d), there are two nanotubes forming the conduction of the transistor we study in this paper. Transistor curve of the device is typical for a quasi-metallic CNT with a small band gap \cite{Amer2013}. Since approximately one third of the nanotubes belong to that type it is most probable that only one of the two CNTs seen in the SEM image contribute to the conductance of the channel at least for positive gate voltages. Nevertheless such an assumption is not crucial for the further analysis of obtained data.  Silicon chip with THz radiation detector in field effect transistor configuration was anchored on an ad hoc ceramic holder. Antennas were coupled with holder contact pads by thin (25~$\mu$m) aluminum wire with help of ultrasonic bonder. Holder was placed in a flow cryostat variable temperature inset. Measurements were made at a temperature of about 100~K. To obtain signals for $p-$ and $n$-type free carriers and controllable vary the carrier density we used a Si back gate applying voltage $V_g$ between Si doped substrate and drain.

The experimental set-up is sketched in Fig.~\ref{characterization}(a). We used normally incident polarized THz radiation of a continuous wave methanol laser operating at a frequency  $f = 2.54$~THz (wavelength $\lambda = 118~\mu$m) with a power $P\approx 20$~mW, see Refs.~\cite{Olbrich2016, Dantscher2017}. The laser spot with a diameter of about 1.5~mm was substantially larger than the device size ensuring uniform illumination of both antennas.  The samples were placed in an optical cryostat and photoresponse was measured as the voltage drop $U$ directly over the sample over the load resistance $R_L =$~1~MOhm applying lock-in technique at a modulation frequency of 77~Hz. Note that due to unknown antenna efficiency we normalize the response U on the total radiation power $P$. The radiation polarization state was controlled by $x$-cut crystal quartz lambda-quarter and lambda-half plates. Rotation of this plates converted initially linearly polarized laser radiation into elliptically one or rotated the plane of incidence by the azimuth angle alpha, respectively. In the former case, rotation of the lambda-quarter plate by and angle $\varphi$ changed the radiation Stokes parameters describing the radiation helicity $P_{\rm circ}$ and the degrees of the linear polarization $P_{L1}$ and $P_{L2}$ after~\cite{Belkov2005,Weber2008}

\begin{equation}\label{circ}
P_{\rm circ} = \sin 2\varphi\:, P_{\rm L1}(\varphi) = \frac{\cos 4\varphi + 1}{2} \:,  P_{\rm L2}(\varphi) = \sin 4\varphi/2\,.
\end{equation}

\section{Results}

\begin{figure}[htbp]
	\centering\includegraphics[width=7cm]{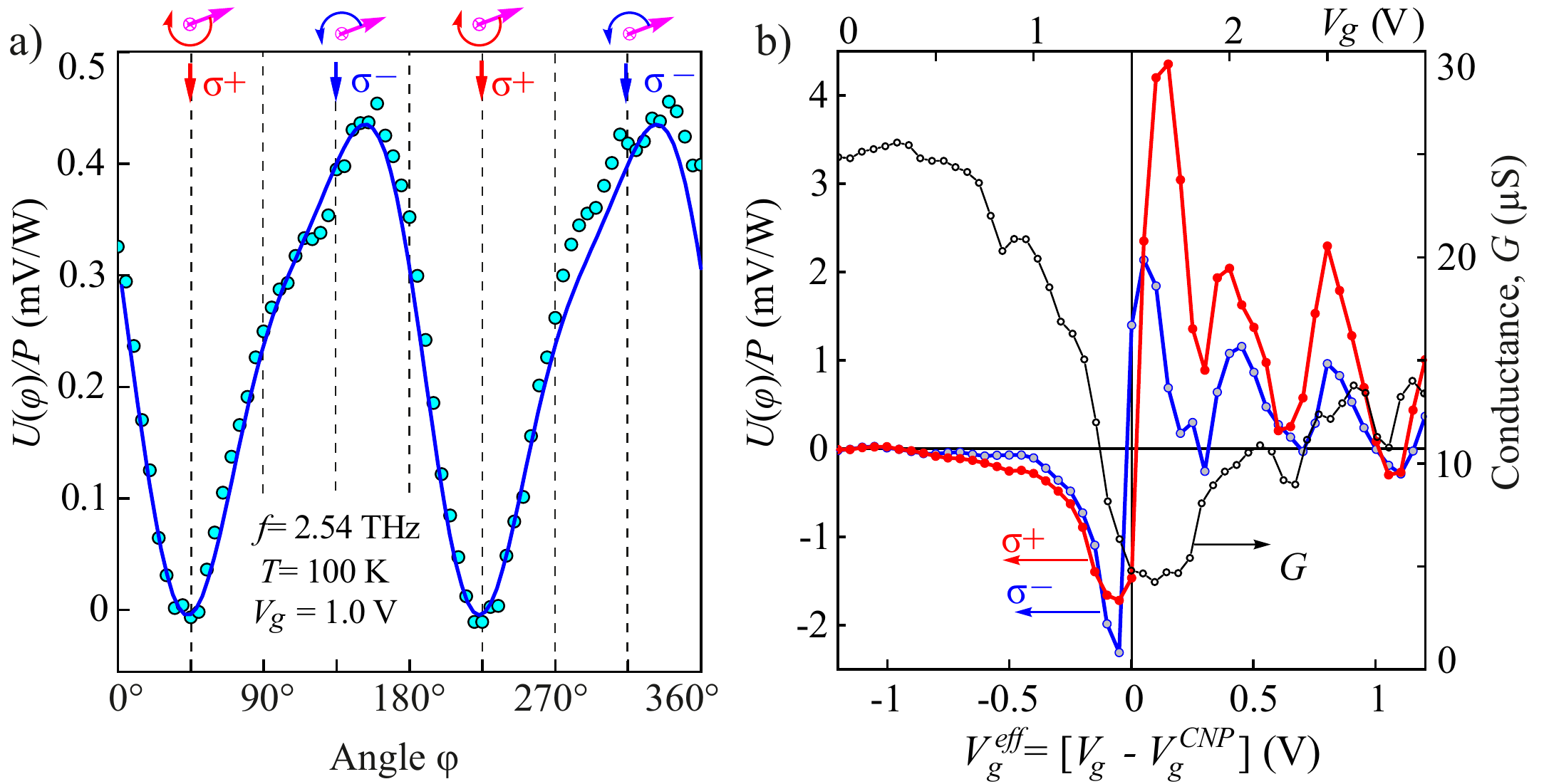}
	\caption{ (a) Helicity dependence of the photovoltage $U(\varphi)$  normalized by the laser radiation power $P$. Downwarded arrows labeled $\sigma^+$ and $\sigma^-$ indicate angles $\varphi$ corresponding to clockwise and anticlockwise rotating in-plane electric field ${\bm E}(t)$ of the right- and left-handed circularly polarized radiation, respectively. Curve is fit after Eq.~(\ref{phi1}).	The ellipses on top illustrate the rotation of the electric field vector for $\sigma^+$ ($\varphi = 45$ and $225$ degrees) and $\sigma^-$ ($\varphi = 135$ and $315$ degrees). (b) Dependencies of the photoresponse $U(\varphi)$	and the sample conductance $G$ on the effective gate voltage $V_g^{\rm eff} = V_g - V_g^{\rm CNP}$, where $V_g^{\rm CNP}$ is a gate voltage corresponding to the charge neutrality point. Upper scale shows the corresponding values of the applied gate voltage $V_g$. Red and blue curves show responses to right- ($U_{\sigma^+}$) and left- ($U_{\sigma^-}$) handed circularly polarized radiation, respectively (see the $y$-axis on the left). Dashed black curve shows the sample conductance measured in the absence of THz radiation ($y$-axis on the right).
	}
	\label{fig:helicity}
\end{figure}

\begin{figure}[htbp]
	\centering\includegraphics[width=7cm]{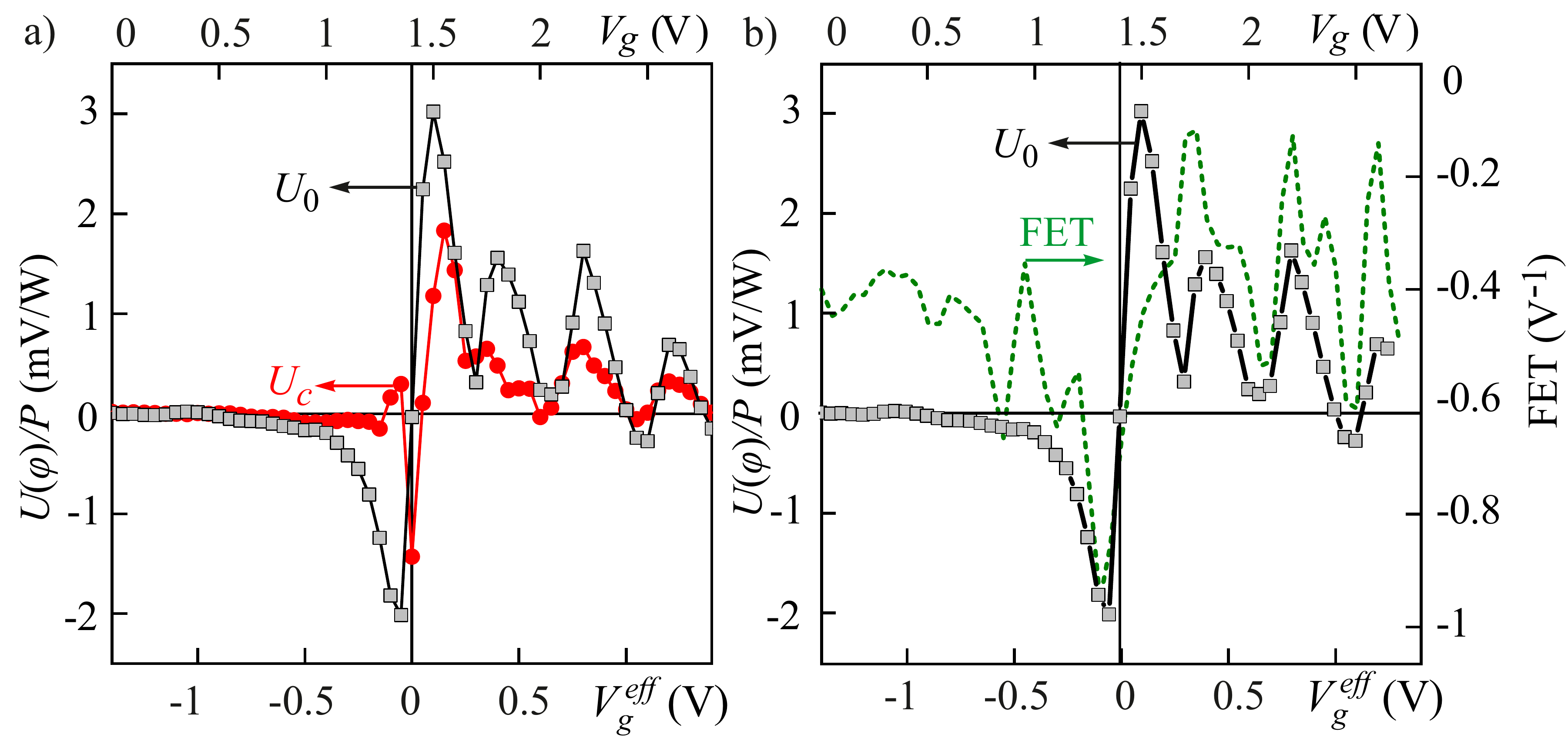}
	\caption{(a) Gate voltage dependencies of the circular ($U_c$) and polarization insensitive ($U_0$) photoresponse normalized on the laser laser power $P$. Taken into account that for the circularly polarized radiation the helicity sensitive contribution changes its sign, whereas $U_0$ remains unchanged and $U_{L1} = U_{L2} = 0$ (both degrees of linear polarization are zero) we calculated the amplitude after $U_{\rm C} = (U_{\sigma^+}-U_{\sigma^-})/2$ and  $U_{\rm 0} = (U_{\sigma^+}+U_{\sigma^-})/2$. Upper scale indicates the corresponding effective gate voltages $V_g^{\rm eff}$. Vertical line indicates the position of the CNP ($U_g^{\rm eff}=0$). (b) Gate dependence of the polarization insensitive response $U_0$ versus the one of the FET factor calculated after Eq.~(\ref{FET}).
	}
	\label{fig:responses}	
\end{figure}

While irradiating the device by normally incident circularly polarized radiation a photoresponse has been observed in unbiased samples. The photocurrent changes significantly if the circular polarization is switched from right-handed to left-handed circularly polarized radiation. Varying the radiation polarization state by rotation of the lambda-quarter plate we detected that the $\varphi$ dependence of the signal is well described by

\begin{equation}\label{phi1}
U = U_c P_{circ} + U_{L1} P_{L1} + U_{L2}P_{L2} + U_0 \,,
\end{equation}
where $U_c$, $U_{L1}$,  $U_{L2}$ and $U_0$ are fitting parameters describing photoresponse contributions sensitive to the radiation helicity ($P_c$) and degrees of linear polarization ($P_{L1}$ and $P_{L2}$) as well as the polarization insensitive one ($U_0$). The dependence is examplary shown for $V_g = 1$~V in Fig.~\ref{fig:helicity}(a). In case of circulalry polarized light the second and thrid terms in the equation (\ref{phi1}) turn to zero. This, together with the fact that only the first term changes the sign by reversing the radiation helicity, makes the fitting procedure straightforward. Varying the effective gate voltage $V^{\rm eff}_{g} = V_g - V^{\rm CNP}_{g}$ results in a complex behavior of the photoresponse. $V_g^{\rm CNP}$ is defined as the gate voltage at which the conductance of the device is minimal. We associate it with the charge neutrality point (CNP) Figure~\ref{fig:helicity}(b) shows the dependencies obtained for right- and left- handed circularly polarized radiation. The figure reveals that the response  changes its sign at the CNP, which can naturally be attributed to the change of carrier type from positively charged holes to negatively charged electrons, and exhibits oscillations for positive gate voltages. In Fig.~\ref{fig:responses}(a)  we plotted values of the helicity sensitive ($U_c$) and polarization insensitive ($U_0$) contributions calculated based on the exeprimental data. These curves reveal that while at positive effective gate voltages both contributions show similar oscillatory behavior, at small negative effective gate voltages the circular response show clear oscillations, whereas
both, $U_0$-contribution and the conductance $G$, shows smooth dependences on the gate voltage.

It is well known that the broad band rectification of terahertz radiation in an FET channel is proportional to the so-called FET  factor
\begin{equation}\label{FET}
{\rm FET} = G^{-1} \frac{dG}{dV_G}.
\end{equation}
Therefore, in Fig.~\ref{fig:responses}(b)  we compared the circular response with the FET factor. We note that the $U_C(V_g)$ and $FET(V_g)$ curves have similar shapes. Both values turn to zero at the CNP and change sign at this point. Moreover, in both cases there is a decreasing trend as the gate voltage goes away from the CNP. We also note oscillations observed for $V_g > V_g^{\rm CNP}$. It is important to note that the oscillations in the $U_C(V_g)$ curve are not directly related to those in the $FET_C(V_G)$ dependence. The minima of conductance occuring to the right from the CNP may be due to irregularities of the density of states at the Fermi level \cite{Wang2005} in an incommensurate DWNT. They should clearly somehow affect the photoresponse of our device. Still detailed analysis and understanding of these conductance and photoresposne oscillations are far beyond of the scope of this paper. This effect has  recently been detected in twisted layer graphene close to the first magical angle~\cite{Otteneder2020}.

Finally, we compared the gate dependence of the conductance measured in the dark and under irradiation with cw terahertz laser radiation (not shown). The obtained results reveal that in the described experiments carried out at $T \approx 100$~K the differential conductance is not affected by the radiation, meaning that the bolometric effect~\cite{Santavicca2011Kim,Chudow2012Kim} is not significant.

\section{Theory}

Below we present theoretical calculation of the interference  contribution to dc response based on approach used before for analysis of helicity-sensitive effects in 2D systems \cite{Gorbenko2018,Gorbenko2019,Matyushkin2020} with   minor modifications for quasi-1D and 1D cases. We restrict ourselves to analysis of the interference helicity-sensitive contribution to  dc response far from the neutrality point. We start with hydrodynamic equations describing  nanotube channel far from the neutrality point, say, at the electron side of the spectrum:
\begin{align}
	&\frac{\p n}{\p t}+\frac{\p v}{\p x}+\frac{\p (nv)}{\p x}=0,
	\label{dndt}
	\\
	&\frac{\p v}{\p t}+ a v \frac{\p v}{\p x}+\gamma(n) v=-s^2(n)\frac{\p n}{\p x}.
	\label{dvdt}
\end{align}
Here $v$ is the drift velocity,  $n=(N-N_F)/N_F$ is the dimensionless concentration $N=N(x,t)$ is the concentration in the channel per unit lenght, $N_F$ is the equilibrium concentration, $a$ is a coefficient on the order of unity, which we discuss below, $s^2(n)\approx s_F^2 (1- \kappa n) $, $\gamma (n)\approx\gamma_F (1+ \xi n)$ are the plasma wave velocity and damping due to the momentum relaxation, respectively. Both $s(n)$ and $\gamma(n)$ depend on the electron concentration (this dependence was neglected in Ref.~\cite{Matyushkin2020}, where plasmonic interference was studied in 2D graphene sample).
Here $s_F$ and $\gamma_F$ are  their values at the Fermi level and terms $-\kappa n$ and $\xi n$  represent first order corrections with respect to $n$. The main difference of  Eqs.~\eqref{dndt} and \eqref{dvdt} as compared to equations used in Refs.~\cite{Gorbenko2018,Gorbenko2019} for 2D electron gas
with parabolic spectrum is related to different type of nonlinearities. First of all, plasma
wave velocity  depends on the electron concentration. Physically, this can be explained as
follows. Electric force $F$ acting on electron in the  channel of the nanotube with inhomogeneous concentration  leads to  the following contribution  to the  electron acceleration $\p v/\p t$
\[\frac{F}{m(n)}=- \frac{e^2 \Lambda
}{m(n)} \frac{\p N}{\p x} \approx  s_F^2(1- \kappa n) \frac{\p n}{\p x} ,  \] where
$m(n)=E(n) /v_F^2 \approx m_F (1+\kappa n)$ is so called fictitious mass, which depends on local electron energy, $E,$ and cosequently, on the local electron concentration,      $s_F^2=e^2 N_F\ln(d/a)/m_F$ is the  squared plasma wave velocity for electrons at the Fermi level \cite{Ryzhii2016} with  $\ln(d/a)$ appearing  due to $1D$ electrostatics of the problem  ($a$ and  $d$  are  nanotube  width and the distance to the  gate, respectively).  The correction to this velocity appears  due to dependence of $m$ on $n.$
We also  take into account  dependence of the scattering rate on the energy and, consequently, on the electron concentration. The  dimensionless factors
$\kappa=m_F^{-1}({\p m}/{\p n})_{n=0} ,$
 $\xi = \gamma_F^{-1}(\p \gamma/\p n)_{n=0}   $ depend
on the type of the nanotube (singlewall or multiwall). Also, factor $\xi$ depends  on the type of the disorder potential.

Dependencies of $s$ and $\gamma$ on $n$ represent two types of nonlinearities,
which have not been discussed in Refs.~\cite{Gorbenko2018,Gorbenko2019,Matyushkin2020}.

 The hydrodynamical  term $a v \p v/\p x$ deserves special  attention. In $2D$ system  with parabolic spectrum, $a=1.$ However, this is not the case for systems with linear spectrum.
 Moreover, $a$ is different for a 2D system with linear spectrum, quasi-1D system, like multi-wall nanotube, and 1D system,  like single-wall nanotube.  As was very recently  shown in Ref.~\cite{Aizin}, in 2D system with linear spectrum,  $a=0.$  Specifically, in such system, the  term, analogous to $v \p v/\p x$  for $v\ll v_F $ looks like
$ (v^3/2V_F^2) dv/dx$ [see Eq.~(17) in Ref.~\cite{Aizin}] and therefore can be neglected within linear-in-radiation power approximation.  In this work, we model a  quasi-1D
multi-wall nanotube as a narrow strip of 2D material, i.e. assume that $a=0$ for this case.
In a full analogy with   calculations of  Ref.~\cite{Aizin}, one  can find for purely 1D case: $a=-1$ (we do not discuss here Luttinger  liquid effects, considering  nanotube
within simplest  semiclassical  approximation based on the Boltzmann equation).

In order to compare results with the ones presented in Refs.~\cite{Gorbenko2018,Gorbenko2019} for parabolic spectrum, we present all equations assuming presence of term $ a (vdv/dx)$ in   Eq.~\eqref{dvdt}  ($a=1$ for  2D system with parabolic spectrum, $a=0$ for multi wall nanotube, and  $a=-1$ for a single wall nanotube)  and for arbitrary values of $ a,$ $\kappa$ and $\xi.$

We assume that two antennas are attached to our device, so that  voltages and concentrations
at source and drain oscillate with the frequency  of incoming radiation $\omega.$ Due to
asymmetry  of the    device a phase shift $\theta$ appears between these signals.
In a full analogy with Refs.~\cite{Gorbenko2018,Gorbenko2019,Matyushkin2020} we use for 
the following  boundary conditions for oscillating part of voltage and concentration,
$n_{ac}(x=0)= U_a \cos (\omega t +\theta) /U_g$, $n_{ac}(x=L)= U_b \cos (\omega t )  /U_g.$      Here $U_a$ and  $U_b$ are the amplitudes of two phase-shifted (by phase $\theta$) ac signals applied to the source and drain, respectively, via two antennas, and $U_g$ is the gate potential, which controls $N_F, ~m_F$  and $s_F.$    General formulas for response are presented in~\hyperlink{General}{Appendix}. Here we only  present expression for interference contribution to the dc response and its asymptotic in the non-resonant case realized in our experiment
\begin{align}
    U_{\rm int}&= \frac{C U_a U_b  }{4U_{g} }~\frac{   \omega\sin\left(\frac{\Omega L}{s_{F}}\right)
   \sinh\left(\frac{\Gamma L}{s_{F}}\right) \sin \theta } { \sqrt{\omega^2+\gamma_F^2}|\sin(kL)|^2}
\label{Uint1}
   \\
   &\approx \frac{ C U_{ a} U_{ b} e^{-L/L_*} \sin(L/L_*) (\omega/\gamma) \sin \theta } {U_{\rm g}}.
\label{Uint2}
\end{align}
where, $k=(\Omega+  i\Gamma)/s_F$ is the complex wave vector of the plasma wave,
and $\Omega, \Gamma$ obey $\Omega+  i\Gamma=\sqrt{\omega(\omega+ i \gamma)}.$
One of the main conclusion  is that this term is not very sensitive to the type of nonlinearity
in the problem. Different nonlinear effects only change numerical coefficient
\be
C=4\left(1+a-\xi\right),
\ee
entering Eqs.~\eqref{Uint1} and \eqref{Uint2} (we notice that $\kappa$  does not enter this equation).

This implies some universality of helicity-sensitive interference response: all details of the spectrum and scattering rate are fully incorporated into a single numerical coefficient which does not depend neither from radiation frequency nor from the fundamental resonant frequency of the device $\omega_0=\pi s_F/L.$  We do not focus here on exact   calculation of $C,$ which for the case  of a single wall nanotube implies analysis of the Luttinger liqud effects encoded in the factor $\xi.$ Instead, we use $C$ as phenomenological parameter on equal footing with $U_a$ and $U_b.$

 Equation \eqref{Uint1} represents    general formula for the interference contribution  valid  for arbitrary relation between radiation frequency,  plasmonic frequency $\omega_0=\pi s_F/L,$ and plasmonic damping $\gamma_F.$
This equation  simplifies   to  \eqref{Uint2} in the non-resonant  regime (and sufficiently long sample) realized in our experiment.In this case,

$\omega <
\gamma_F,~\omega_0 < \gamma_F ,$ and  $L >
L_*$
where,
\begin{equation}
	L_*= \frac{s_F\sqrt 2}{\sqrt {\omega \gamma_F}}
	\label{eqn:Lstar}
\end{equation}
is the plasma wave decay length, which we estimate for our experiment as $L_* \approx 0.2 ~\mu$m.

In this regime, plasma waves,
excited at the source and drain parts of the channel weakly overlap
inside the transistor  channel.

As was shown in Ref.~\cite{Matyushkin2020}, the response can be easily rewritten in terms of
the Stokes parameters if we model two antennas attached to the source and drain by  thin
metallic rods of lengths $R_a$ and $R_b,$ respectively, rotated by angles $\theta_a$ and $\theta_b$ in the plane of the device. The circular  contribution to  response  is  equal to  the interference term  and  is given by
\be
\begin{aligned}
	\label{interference}
	&U_{\rm circ} = U_{\rm int} =\frac{U(\phi = 45^\circ) -U(\phi = 135^\circ)}{2}
	\\
	&= -\frac{C E_0^2 R_{ a } R_{b}}{ 4 U_{\rm g}}
 \frac{\omega}{\gamma} \sin(L/L_*) \exp(-L/L_*)\sin(\theta_a-\theta_b),
\end{aligned}
\ee
where $E_0$ is the amplitude of the cicrcularly polarized wave. Since Eq.~\eqref{interference}
contains factor $\exp(-L/L_*),$  the main contribution to gate voltage dependence
of the response is due to dependence  $L_*$ on $U_g.$ 

\section{Discussion of the experimental and theoretical results}

The most important experimental result of this work is the observation of the helicity dependent photoresponse of the CNT based structure, see Figs. \ref{fig:helicity} and	\ref{fig:responses}. So far, such a photocurrent in CNT was considered theoretically in Ref.~\cite{Ivchenko2002pgeCNT} demonstrating that a response proportional to the degree of circular polarization $P_{circ}$ is expected for chiral nanotubes. This mechanism, however, cannot be responsible for the observed effect. Indeed, it can only be excited by radiation propagating along the CNT axis~\cite{Ivchenko2002pgeCNT}. Consequently, in our experiments, applying normally incident radiation to the nanotube lying on the structure surface, it is forbidden by the symmetry arguments. Moreover, an essential part of the structure are antenna sleeves used for the radiation coupling.  For elliptical (circular) radiation the antennas highly amplify a particularly oriented radiation electric field vector $\bm E$ so that a photoresponse is, in fact, excited by the linear polarized radiation. Therefore, the helicity dependent signal may only results from the shifted in time excitation of the source and drain sleeves, with the phase shift different for the clockwise and anti-clockwise rotating radiation $\bm E$, i.e. right- and left-handed circularly polarized radiation. This mechanism considered theoretically in Refs.~\cite{Drexler2012,Romanov2013,Gorbenko2018,Gorbenko2019} was most recently realized in graphene structures~\cite{Matyushkin2020} and is caused by plasmon interference.

Now we turn to the gate dependence of the  photoresponse. We begin with the sign inversion observed close to CNP. As addressed above, it is caused by the change of sign of the carrier type, and, consequently, sign inversion of the photoresponse. The photovoltage here is  
 proportional to the first derivate of the conductance in Eq.~\ref{FET} well established for the plasmonic response. 

Let us compare experimental results with Eq.~\eqref{interference}. In contrast to our previous estimates for $2D$
case \cite{Matyushkin2020}, in 1D case both for short range of charged impurities the dependence
of $\gamma_F$ on energy is quite slow, and one can neglect gate voltage dependence of $L_*.$ This
conclusion agrees
with our observations.  Indeed, as  we above discussed above,     interference part of the response
decays with $U_g$ approximately (see, Fig.~\ref{fig:responses}) as  $1/U_g$ is agreement with Eq.~\ref{interference}
 with fixed $L_*$.  The factor $\sin(L/L_*)/U_g$ in the eq. ~\ref{interference} describes oscillations of the  $U_{\rm circ}$ as a function of the gate voltage close to the CNP where the dependence of $L_*$ on the gate voltage cannot be neglected. This may explain the experimental data shown in the fig. 3a. close to $V_g = 0$.

\section{Summary}
In this work,  we investigated photovoltaic   terahertz radiation response of a    carbon nanotube electrically coupled with a gate and  two antennas asymmetrically attached to source and drain.  We demonstrated, both experimentally and theoretically,  that   response contains  a large  gate-controllable contribution  arising due to the interference of plasma wave  propagating from source and drain inside the channel.     
Specifically, we experimentally demonstrated  that the magnitude of the rectified voltage is different for clockwise and anticlockwise circularly polarized radiation, i.e. is sensitive to the phase shift   appearing as a consequence of the asymmetry of the antenna configuration. We developed a theory, which explains basic experimental results.  Importantly,     our  theoretical analysis showed that this  effect is universal  in a sense that the    helicity- and phase-sensitive part of the   response is the same (up to unimportant numerical coefficients)   in  the systems of different dimensionalities   with different single particle spectrum.    For non-resonant  regime, discussed in the current work,  the  magnitude of the response  is  fully controlled by  the plasma wave decay length. 

\section*{Acknowledgments}
The authors acknowledge the support of the DFG-RFFI project (Ga501/18-1, RFFI project 2152-12015), RSF proj. 21-72-20050 (device fabrication and characterization), and the Elite Network of Bavaria (K-NW-2013-247). VK and SG acknowledge the support of the IRAP Programme of the Foundation for Polish Science (grant MAB/2018/9, project CENTERA). The theoretical part of the was funded by the Russian Science Foundation, Grant No. 20-12-00147.

\appendix

\section{Appendix: General formulas for response} 
\label{General}
Calculations, analogous to ones, presented in \cite{Gorbenko2018}, yield the following formula for dc photoresponse
\be
U_{dc}= \frac{\omega}{\sqrt{\omega^2+\gamma_F^2}}\frac{\alpha (U_a^2-U_b^2)+ \beta U_a U_b \sin \theta } {4U_{g}|\sin(kL)|^2}
\label{voltage}
\ee
where
\begin{equation*}
\begin{aligned}
	\alpha &= \left\{\!
	\left[a\! +\! \frac{(1\! -\! \xi)\gamma_F\Omega}{\Gamma\omega}\right]\! -\!
	\kappa \frac{\sqrt{\omega^2\! +\! \gamma_F^2}} {\omega}
	\right\}\!
	\sinh^2\! \left(\! \frac{\Gamma
		L}{s_{F}}\! \right)\! + 
	\\
	&+
	\left\{\! \left[\frac{(1\! -\! \xi)\Gamma\gamma_F}{\Omega\omega}\! -\! a\right]\!
	-\! \kappa  \frac{\sqrt{\omega^2\! +\! \gamma_F^2}} {\omega}\right\}\!\sin^2\!\left( \!\frac{\Omega L}{s_{F}}\!
	\right)\!,
\end{aligned}
\end{equation*}
Here,
\be
\beta= C \sin\left(\frac{\Omega L}{s_{F}}\right)\sinh\left(\frac{\Gamma L}{s_{F}}\right),
\ee
where $k=(\Omega + i \Gamma)/s_F= \sqrt{\omega(\omega+ i \gamma)}/s_F, $ and
$C=4\left(1+a-\xi\right).$ The term $|\sin(kL)|^2$  in denominator of  Eq.~\ref{voltage} is responsible for plasmonic resonance.
As seen, response depends on numerical coefficients  $a,\kappa$ and $\xi.$   The second term
in Eq.~\ref{voltage} is proportional to  $U_aU_b$ and is responsible
for the plasma wave interference effects.
 Different nonlinear effects only change numerical coefficient in front of this
terms (in our case, coefficient $8$ for 2D system  with  parabolic spectrum and short range scattering
transforms to $4(1-\xi)$ for the linear spectrum).


\end{document}